\documentclass[color,french,12pt]{article}
\usepackage{amsmath,amssymb}
\usepackage{fancybox}
\usepackage[latin1]{inputenc}
\usepackage{amsmath,amssymb}
\usepackage{graphicx}
\usepackage{array}
\usepackage{multimedia}
\usepackage[latin1]{inputenc}
\usepackage{euscript}
\usepackage{appendix}
\usepackage{amsthm}
\usepackage{minitoc}
\usepackage{fancyhdr}

\input{tcilatex}
\begin{document}

\title{\textbf{ON HAWKING RADIATION OF \\
$2d$ LIOUVILLE BLACK HOLE}}
\author{M. Chabab$^1$ , H.EL Moumni$^1$, M.B. Sedra$^2$ \and {\small 1 :
LPHEA University of Cadi Ayyad Marrakech, FSSM, Department of physics,
Morocco.} \and {\small 2 : LHESIR, Ibn Tofail University at Kenitra, FSK,
Department of physics, Morocco.}}
\maketitle

\begin{abstract}
We adapt the method of complex paths to the study of the radiation of
Hawking of Liouville black holes.
\end{abstract}


\section{Black hole in $2d$-gravity}

The study of the Liouville black holes is a research topics whose interest
is increasing \cite{1, 2, 3, 4, 5, 6}. In the present note, we start by
considering, as a review, the action of $Einstein$ gravity that we write as
\cite{2, 3}
\begin{equation}
S\left[ g_{\mu \nu },\phi \right] =\int d^{2}x\sqrt{-g}\left[ \frac{1}{16\pi
G}\left( \psi R+\Lambda +\frac{1}{2}(\nabla \psi ^{2})\right) +\mathcal{L}%
_{m}\right] ,  \label{1}
\end{equation}%
where $R$ is the $Ricci$ scalar, $\psi $ is a dilatonic field and $\mathcal{L%
}_{m}$ is the mater Lagrangian. The $G$ and $\Lambda $ are the gravitational
and cosmological constant. The variation with the respect to $\psi $ and $%
g_{\mu \nu }$ give the following equations of motion:
\begin{equation}
R-\nabla ^{2}\psi =0,  \label{2}
\end{equation}%
and
\begin{equation}
\frac{1}{2}\nabla _{\mu }\psi \nabla _{\nu }\psi -g_{\mu \nu }\left( \frac{1%
}{4}\nabla ^{\mu }\psi \nabla _{\nu }\psi \right) -\nabla _{m}u\nabla
_{n}u\psi =8\pi GT_{\mu \nu }+\frac{1}{2}\Lambda g_{\mu \nu },  \label{3}
\end{equation}%
where the stress-energy tensor is
\begin{equation}
T^{\mu \nu }=-\frac{2}{\sqrt{-g}}\frac{\delta \mathcal{L}_{m}}{\delta g_{\mu
\nu }}.  \label{4}
\end{equation}%
By tracing (\ref{2}), the dilaton $\psi $ decouples from Einstein equations,
simplifying (\ref{1}) to the standard Liouville gravity form
\begin{equation}
R-\Lambda =8\pi GT  \label{5}
\end{equation}%
where $T=T_{\mu }^{\mu }$

By prescribing the energy momentum tensor $T^{\mu \nu }$ one gets a variety
of solutions with nontrivial horizon structure. For instance, in the case of
a static source representing a point particle at the origin, the energy
density is given by

\begin{equation}
\rho =\frac{M}{2\pi G}\delta (x)  \label{6}
\end{equation}%
where $M$ corresponds to the ADM mass. \ Note that for the generic line
element

\begin{equation}  \label{7}
ds^2=-\alpha(x)\ dt^2+\alpha^{-1}(x)\ dx^2
\end{equation}
the equations of motion (\ref{5}) reads as

\begin{equation}
\frac{d^{2}}{dx^{2}}\alpha (x)+\Lambda =4M\delta (x).  \label{8}
\end{equation}%
Assuming a linearly-symmetric solution about the origin ($\alpha =\alpha
(|x|)$), this becomes
\begin{equation}
\alpha ^{\prime \prime }+2\alpha ^{\prime }\;\delta (x)+\Lambda =4M\delta
(x)~~,  \label{9}
\end{equation}%
where the prime symbol is nothing but the derivation with respect the
coordinate $x$ namely $d/d|x|$. If $\alpha $ is continuous then one is led
to the consistency condition $2\alpha ^{\prime }=4M$ at $x=0$. It can be
shown \cite{3} that the solution is $\alpha (x)=\frac{1}{2}\Lambda
x^{2}+2M|x|-C$, and so the two-dimensional metric reads as
\begin{equation}
ds^{2}=-\left( -\frac{1}{2}\Lambda x^{2}+2M|x|-C\right) dt^{2}+\frac{dx^{2}}{%
\left( -\frac{1}{2}\Lambda x^{2}+2M|x|-C\right) }.  \label{10}
\end{equation}%
with $C$ an arbitrary constant. The horizon can be described explicitly as
\begin{equation}
|x|_{0}=\frac{2M\pm \sqrt{4M^{2}-2\Lambda C}}{\Lambda },  \label{11}
\end{equation}

\section{Hawking radiation and the method of complex paths}

The first part of this section is a cover of an important formalism,
presented in Section 3 of \cite{1}, that we will exploit in order to extract
some important properties of the Hawking radiation. In this issue, we use
the method of complex path analysis \cite{1}. The line element is given by
equation ($\ref{5}$), $\alpha (x)$ vanishes at $x_{0}$, and $\alpha ^{\prime
}(x)$ is nonzero at $x_{0}$. Expanding $\alpha (x)$ around the point $x_{0}$
gives
\begin{equation}
\alpha (x)=\alpha ^{\prime }(x_{0})(x-x_{0})+\mathcal{O}\left[ (x-x_{0})^{2}%
\right] \equiv \mathcal{R}(x_{0})(x-x_{0})  \label{12}
\end{equation}%
where $\mathcal{R}(x)=2M-x\Lambda $ and $\mathcal{R}(x_{0})=\sqrt{%
4M^{2}-C\Lambda }\neq 0\Rightarrow M^{2}\neq C\Lambda $. Now consider a
scalar field which satisfies the Klein-Gordon equation
\begin{equation}
\left( \square -\frac{m_{0}^{2}}{\hbar ^{2}}\right) \Phi =0  \label{13}
\end{equation}%
In the background of $\ref{7}$ the last equation can be written as follows
\begin{equation}
-\frac{1}{\alpha (x)}\frac{\partial ^{2}\phi }{\partial t^{2}}+\frac{%
\partial }{\partial x}\left( \alpha (x)\frac{\partial \Phi }{\partial x}%
\right) =\frac{m_{0}^{2}}{\hbar ^{2}}\Phi   \label{14}
\end{equation}%
The semiclassical wave functions satisfying the above are obtained by making
the standard ansatz
\begin{equation}
\phi (x,t)=e^{\frac{i}{\hbar }\mathcal{S}(x,t)}  \label{15}
\end{equation}%
Substituting this ansatz into equation($\ref{14}$) gives
\begin{equation}
-\frac{1}{\alpha (x)}\left( \frac{\partial \mathcal{S}}{\partial t}\right)
^{2}+\alpha (x)\left( \frac{\partial \mathcal{S}}{\partial x}\right)
^{2}+m_{0}^{2}-\frac{i}{\hbar }\left[ \frac{1}{\alpha (x)}\frac{\partial ^{2}%
\mathcal{S}}{\partial t^{2}}-\alpha (x)\frac{\partial ^{2}\mathcal{S}}{%
\partial x^{2}}-\frac{d\alpha (x)}{dx}\frac{\partial \mathcal{S}}{\partial x}%
\right]   \label{16}
\end{equation}%
Now we expand $\mathcal{S}$ in a power series of $\hbar /i$
\begin{equation}
\mathcal{S}(x,t)=\sum_{n=0}\left( \frac{\hbar }{i}\right) ^{n}\mathcal{S}%
_{n}(x,t)  \label{17}
\end{equation}%
Neglecting terms of hight order in $\hbar /i$ $\mathcal{S}_{0}$ gives rise
to
\begin{equation}
-\frac{1}{\alpha (x)}\left( \frac{\partial \mathcal{S}_{0}}{\partial t}%
\right) ^{2}+\alpha (x)\left( \frac{\partial \mathcal{S}_{0}}{\partial x}%
\right) ^{2}+m_{0}^{2}=0  \label{18}
\end{equation}%
and the solution is given by
\begin{equation}
\mathcal{S}_{0}=-\mathcal{E}t\pm \int^{r}\frac{dx}{\alpha (x)}\sqrt{\mathcal{%
E}^{2}-m_{0}^{2}\alpha (x)}  \label{18}
\end{equation}%
$\mathcal{E}$ is a constant witch is identified to energy. To simplify, we
take $m_{0}=0.$ Using the usual saddle point method, the semiclassical
propagator $\mathcal{K}(z^{\prime \prime },z^{\prime })$ for a particle
propagating from a spacetime point $z^{\prime \prime }(t_{1},x_{1})$ to $%
z^{\prime }(t_{2},x_{2})$ is given by
\begin{equation}
\mathcal{K}(z^{\prime \prime },z^{\prime })=\mathcal{N}\exp \left[ \frac{i}{%
\hbar }\mathcal{S}_{0}(z^{\prime \prime },z^{\prime })\right]   \label{19}
\end{equation}%
with $\mathcal{N}$ is a normalization constant and $\mathcal{S}_{0}$ is
given by

\begin{equation}
\mathcal{S}_{0}(z^{\prime \prime },z^{\prime })=-\mathcal{E}(t_{2}-t_{1})\pm
\int_{x1}^{x2}\frac{dx}{\alpha (x)}  \label{20}
\end{equation}%
We can compute the amplitudes and probabilities of emission and absorption
through the event horizon at $x_{0}$. Next, consider an outgoing particle at
$x=x_{1}<x_{0}$, the modulus squared of the amplitude for this particle to
cross the horizon gives the probability of emission of the particle.
Invoking the usual $i\epsilon $ prescription, the contribution to $S_{0}$ in
the ranges $(x,x_{0}-\epsilon )$ and $(r_{0}+\epsilon ,r)$ is real. We take
the contour to lie in the upper complex plane and find

\begin{eqnarray}
\mathcal{S}_{0}[emission] &=&-\mathcal{E}\lim_{\epsilon \rightarrow
0}\int_{r_{0}-\epsilon }^{r_{0}+\epsilon }\frac{dx}{\alpha (x)}+\text{real
part}  \label{21} \\
&=&\frac{i\pi \mathcal{E}}{\pm \sqrt{4M^{2}-2C\Lambda }}+\text{real part}
\end{eqnarray}%
and
\begin{eqnarray}
\mathcal{S}_{0}[absorption] &=&-\mathcal{E}\lim_{\epsilon \rightarrow
0}\int_{r_{0}+\epsilon }^{r_{0}-\epsilon }\frac{dx}{\alpha (x)}+\text{real
part}  \label{22} \\
&=&-\frac{i\pi \mathcal{E}}{\pm \sqrt{4M^{2}-2C\Lambda }}+\text{real part}
\end{eqnarray}%
Squaring the modulus to get the probability gives,
\begin{equation}
P[emission]\varpropto \exp \left[ \frac{-2\pi \mathcal{E}}{\pm \hbar \sqrt{%
4M^{2}-2C\Lambda }}\right] \text{ }\text{ }\text{ , }\text{ }\text{ }%
P[absorption]\varpropto \exp \left[ \frac{2\pi \mathcal{E}}{\pm \hbar \sqrt{%
4M^{2}-2C\Lambda }}\right]  \label{23}
\end{equation}%
implying that
\begin{equation}
P[emission]=\exp \left[ \frac{-4\pi \mathcal{E}}{\pm \hbar \sqrt{%
4M^{2}-2C\Lambda }}\right] P[absorption]  \label{24}
\end{equation}%
Comparing this formula with the relation due to Hawking and Hartle,
\begin{equation}
P[emission]=e^{-\beta \mathcal{E}}P[absorption],  \label{25}
\end{equation}
we obtain the standard expression of the Hawking temperature of $(1+1)$
Liouville black hole, namely
\begin{equation}
T_{H}=\beta ^{-1}=\frac{\hbar }{2\pi }\sqrt{M^{2}-C\frac{\Lambda }{2}}
\label{27}
\end{equation}%
A two dimensional version of $Stefan^{\prime }s$ law gives the total power
radiated by the black hole:
\begin{equation}
\mathcal{P}\thicksim -\frac{dM}{dt}\thicksim T_{H}^{2}=\frac{\hbar ^{2}}{%
4\pi ^{2}}\left( M^{2}-C\frac{\Lambda }{2}\right)  \label{28}
\end{equation}%
Using the power $\mathcal{P}$ it is possible to estimate the evaporation
time of the Liouville black hole
\begin{equation}
t_{evap}\thicksim \frac{M}{|dM/dt|}=\frac{4\pi ^{2}}{\hbar ^{2}}\frac{M}{%
\left( M^{2}-C\frac{\Lambda }{2}\right) }  \label{29}
\end{equation}

To conclude, we point out that our principal objective, through this work,
consists in studying the radiation of Hawking in the case of Liouville black
holes. This is done by using the method of complex paths developed in \cite%
{1}, The obtained results, in this work, are similar to those obtained in
\cite{2}. Besides the increasing interest in black holes physics, we have to
underline that the relevance of our work can be related also to the
importance of the complex paths method. We will focus, in the forthcoming
occasion, to push much more this study to other topics in black holes and
string theory.

\end{document}